\documentclass{article}
\usepackage{hq2kpro}
\usepackage{epsf}
\newif\ifpreprint
\preprinttrue

\ifpreprint\usepackage{jurgen}\fi

\begin{document}

\title{
\ifpreprint
\rightline{UASLP--IF--00--05}
\rightline{Fermilab--Conf--00/309--E}
\fi
RECENT RESULTS ON CHARM AND HYPERON PHYSICS FROM
\ifpreprint
SELEX
\else 
SELEX
\fi
}
\author{
\ifpreprint 
\begin{minipage}{\hsize}
\centering
J.~Engelfried$^{m,e}$,
\footnotesize
G.~Alkhazov$^{k}$,
A.G.~Atamantchouk$^{k}$,
M.Y.~Balatz$^{h,\ast}$,
N.F.~Bondar$^{k}$,
P.S.~Cooper$^{e}$,
L.J.~Dauwe$^{q}$,
G.V.~Davidenko$^{h}$,
U.~Dersch$^{i,\dag}$
A.G.~Dolgolenko$^{h}$
% ...
\\ \mbox{~~~}\\ 
% ...
$^e$Fermilab, Batavia, IL 60510, U.S.A.\\
% ...
$^h$Institute of Theoretical and Experimental Physics, Moscow, Russia\\
$^i$Max-Planck-Institut f\"ur Kernphysik, 69117 Heidelberg, Germany\\
% ...
$^k$Petersburg Nuclear Physics Institute, St.\ Petersburg, Russia\\
% ...
$^m$Universidad Aut\'onoma de San Luis Potos\'{\i}, San Luis Potos\'{\i}, Mexico\\
% ...
$^q$University of Michigan--Flint, Flint, MI 48502, U.S.A.\\
% ...
\end{minipage}
\else
J\"urgen Engelfried \\
{\em Universidad Aut\'onoma de San Luis Potos\'{\i}, Mexico} \\
on behalf of the SELEX Collaboration\\ 
\fi
}
\maketitle
\ifpreprint\newpage\fi
\baselineskip=11.6pt
\begin{abstract}
The 
\ifpreprint SELEX \else SELEX\cite{selex} \fi 
experiment (Fermilab E781) is a 3-stage magnetic spectrometer
for the study of charm hadroproduction at large $x_F$ using 
$600\,\mbox{GeV}/c$ $\Sigma^-$, $\pi^-$, and $p$ beams. New precise 
measurements of the $\Lambda_c$, $D^0$, and $D_s$ lifetimes are presented.
Results on $\Lambda_c$ and $D_s$ production for $x_F>0.2$ are reported
as well.  The spectrometer was also used for hyperon physics, where we
will show measurements of the $\Sigma^-$ charge radius, the polarization of
inclusive produced $\Lambda$s, and the polarization of beam $\Sigma^+$.
\end{abstract}
\baselineskip=14pt
\section{The SELEX Experiment}
The SELEX experiment at Fermilab is a 3-stage magnetic 
spectrometer\cite{spec}.  
The negative beam ($600\,\mbox{GeV}/c$)
had about equal fluxes of $\pi^-$ and $\Sigma^-$.  The positive beam 
($540\,\mbox{GeV}/c$) was $92\,\mbox{\%}$ protons. 
The primary and secondary vertex resolution for typical charm events is
$\sigma_p=270\,\mu\mbox{m}$ and $\sigma_s=560\,\mu\mbox{m}$, respectively.
The momentum resolution for charged tracks is
$\delta p/p\approx 0.5\,\mbox{\%}$ at $100-300\,\mbox{GeV}/c$.
A RICH detector\cite{rich} labelled all particles above $25\,\mbox{GeV/}c$
with high efficiency, greatly reducing background in charm analyses.

Our charm analysis, a vertex driven analysis with definite RICH identification
for all kaon and proton candidates required:
(i)~a good secondary vertex ($\chi^2/dof <5$);
(ii)~a longitudinal separation ($L$) between 
  the vertices bigger than $8\,\sigma$, were $\sigma$ is defined as
 $\sigma^2 = \sigma^2_p + \sigma^2_s$;
(iii)~the reconstructed momentum vector of the charm track to point back 
towards the primary vertex within errors;
(iv)~$K$, $p$ identified by the RICH with
${\cal L}(K)>{\cal L}(\pi)$, ${\cal L}(p) > {\cal L}(\pi)$;
(v)~no secondary vertices inside downstream material.

The mass resolution, 
measured via the width of $D^0\to K^\pm\pi^\mp$
and $D^0\to K^\pm\pi^\mp\pi^+\pi^-$, is constant 
($\sim 9\,\mbox{MeV/}c^2$) over the whole momentum
range of interest.

The geometrical acceptance and detection efficiency
is identical for particle and anti-particle to
$<3\,\mbox{\%}$, still high at large $x_F$, and is flat in $p_T$.

\section{Weak Decays of Charm}
\subsection{Motivation}
Charm lifetime measurements test models 
based on $1/M_{Q}$ QCD expansions and evaluate corrections from 
non-spectator $W$-annihilation and Pauli interference to perturbative QCD 
matrix elements. The lifetime hierarchy for
the charm system presents a challenge to HQET and pQCD methodologies due
to the low charm quark mass.

\subsection{SELEX Lifetime Analysis Method}
SELEX charm signals are extracted by the sideband subtraction method
with a fixed signal region ($\pm 2.5\,\sigma_M$, i.e.\ $20\,\mbox{MeV/}c^2$) 
centered on the 
charm mass.  Sidebands of equal width are defined above and below
the charm mass region.  The background under the charm peak is the average
of the two sideband regions.

$\pi$ and $K$ misidentification causes mixing of charm signals.  In SELEX
this is significant only for the $D_{s}$ peak, due to the excellent particle
identification.  
For both $D_s$ modes kaon momenta are $\le 160\,\mbox{Gev/}c$ to reduce 
misidentification.  
Any $KK\pi$ event having a pseudo-$D^{\pm}$ mass in the
interval $(1867\pm 20)\,\mbox{Mev/}c^2$ is removed to eliminate an 
artificial lengthening of the $D_s$ lifetime, even though some 
of these are real $D_s$ events.

Due to the excellent proper time resolution of $\approx 20\,\mbox{fs}$, we 
make a binned maximum 
likelihood fit simultaneously to signal
and sideband regions in reduced proper time  
$t^* = M(L-8 \sigma)/pc $.

The acceptance correction functions were established using data only, 
which greatly reduces systematic effects.  Overall systematics errors were
studied with different decays modes and by comparing the results with events
from different production targets. The $D^0$ and $D^+$ lifetimes were 
measured mostly to validate
the other measurements, presenting the largest acceptance correction.

\subsection{Results for $\Lambda_c^+$, $D^0$, $D^+$, $D_s$}
\begin{figure}[htb]
\begin{center}
\leavevmode
\epsfxsize=\hsize
\epsffile{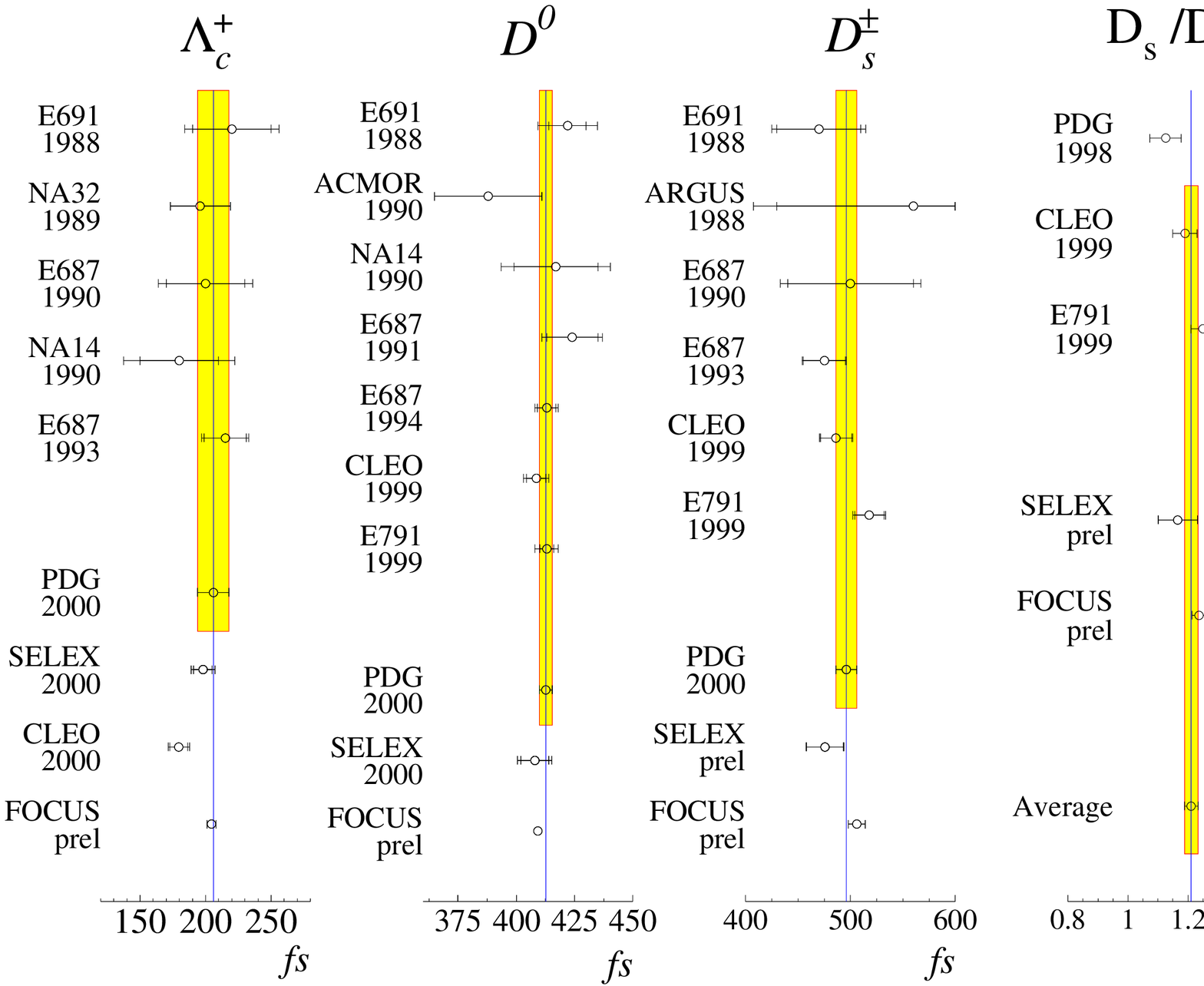}
\caption{\it Summary of recent lifetime measurements for 
$\Lambda_c$, $D^0$, and
$D_s$. Also shown are results for the ratio of $D_s$ to $D^0$ lifetimes.
Lifetime results are taken 
from references\cite{lambdac} to\cite{pdg1998}.}
\label{lifesum}
\end{center}
\end{figure}

The results presented here for the lifetimes of $\Lambda_c^+$, $D^0$, and $D^+$
are submitted for publication\cite{lambdac}, the results for $D_s$ are still
preliminary. A summary is presented in table~\ref{lifetime}.
\begin{table}[htb]
\begin{center}
\caption{ \it SELEX charm lifetime measurements. The results for $D_s$ are
preliminary.
\label{lifetime}
}
\vskip 0.1 in
\begin{tabular}{|l|r|r|r|}
\hline
        &Lifetime & Stat.\ Err.\ & Sys.\ Err.\  \\ \hline
 \boldmath $\Lambda_c^+ \to pK^-\pi^+$ &
 \boldmath $198.1\,\mbox{\bf fs}\;$ & \boldmath  $7.0\,\mbox{\bf fs}\;$ & 
 \boldmath $5.6\,\mbox{\bf fs}\;\;$ \\ &&& \\ \hline
$D^0\to K\pi$ &
$416\,\mbox{fs}\;$ & $8\,\mbox{fs}\;$ & \multicolumn{1}{c|}{--} \\  
$D^0\to K\pi\pi\pi$ &
$402\,\mbox{fs}\;$ & $10.5\,\mbox{fs}\;$ & \multicolumn{1}{c|}{--} \\  
\boldmath $D^0$ &
 \boldmath $407.9\,\mbox{\bf fs}\;$ & \boldmath $6.0\,\mbox{\bf fs}\;$ & 
 \boldmath $4.3\,\mbox{\bf fs}\;\;$ \\  \hline
\boldmath $D^\pm \to K\pi\pi$ &
 \boldmath $1070\,\mbox{\bf fs}\;$ & \boldmath $36\,\mbox{\bf fs}\;$ &
 \multicolumn{1}{c|}{--} \\ \hline
$D_s\to\phi\pi$ &
$474\,\mbox{fs}\;$ & $22\,\mbox{fs}\;$ & \multicolumn{1}{c|}{--} \\ 
$D_s\to K^\star K$ &
$478\,\mbox{fs}\;$ & $33\,\mbox{fs}\;$ & \multicolumn{1}{c|}{--} \\
\boldmath $D_s$&
\boldmath $475.6\,\mbox{\bf fs}\;$ & \boldmath $17.5\,\mbox{\bf fs}\;$ & 
\boldmath $4.4\,\mbox{\bf fs}\;\;$ \\ \hline
\end{tabular}
\end{center}
\end{table}
A summary of recent lifetime measurements for $\Lambda_c$, $D^0$, and
$D_s$ is shown in fig.~\ref{lifesum}.
As can be seen, the SELEX result for the $\Lambda_c^+$ lifetime has a smaller
error than the PDG2000 average.  The result for $D^0$ and $D_s$ is comparable
with previous measurements, but we are looking forward to the new results
from FOCUS.  For the ratio of the lifetimes of $D_s$ and $D^0$, the 
preliminary SELEX result gives $\tau_{D_s}/\tau_{D^0}= 1.166\pm0.065$, 
about 3 standard deviations away from 1.  Taking into account all recent 
measurements, the result is $\tau_{D_s}/\tau_{D^0}= 1.21\pm0.02$, which 
support the necessity\cite{bigi} 
of the $W$-annihilation diagrams in weak decays.

\section{Hadroproduction of Charm}
\subsection{Motivation}
Charm Hadroproduction is a major challenge to factorized perturbative QCD
analysis.
The quark level processes are charm--anticharm symmetric, and next to leading
order terms only introduce a small asymmetry.  
In contrast, some experiments observed large production asymmetries 
in some cases.

\looseness=-1
Two basic models exist which try to explain these asymmetries: the color
drag model, sometimes also called ``leading particle effect''\cite{leading}, 
and the intrinsic charm model\cite{brodski}.
The models differ in the predicted behavior of the asymmetries as functions
of $x_F$ and $p_T$.
SELEX, with its 3~different hadron beams 
($\Sigma^-$, $\pi^-$, $p$), has the unique opportunity to add experimental
data to this problem.

\subsection{SELEX Features}
As mentioned before, SELEX has (by design) a high acceptance at large $x_F$,
the acceptance is constant over a wide range of $p_T$, and is identical
for particle and anti-particle to better than $3\,\mbox{\%}$ in any bin. 
For the asymmetry data, no acceptance correction has been applied.

\subsection{Production of $\Lambda_c$ by $\Sigma^-$, $\pi^-$, and $p$}
SELEX measured the differential production cross sections 
of $\Lambda_c^+$ and $\Lambda_c^-$ with the 3~different beams
as a function of $x_F$ and $p_T$.  The preliminary results for a fit
to $d\sigma/dx_F \sim (1-x_F)^n$ are shown in table~\ref{lambdacn}.
\begin{table}[htb]
\begin{center}
\caption{ \it Preliminary results for a fit to the differential 
cross section ~~~~
$d\sigma/dx_F \sim (1-x_F)^n$ for $\Lambda_c^+$ and $\Lambda_c^-$
for different beams.
\label{lambdacn}
}
\vskip 0.1 in
\begin{tabular}{|l|c|c|}
\hline
Beam & Particle & $n$ \\ \hline
$\Sigma^-$ & $\Lambda_c^+$            & $2.45\pm0.18$ \\
$\Sigma^-$ & $\Lambda_c^-$ & $6.8\pm 1.1 $\\ \hline
$\pi^-$    & $\Lambda_c^+$            & $2.65\pm0.44 $\\
$\pi^-$    & $\Lambda_c^-$ & $2.2\pm0.8 $\\ \hline
$p$        & $\Lambda_c^+$            & $2.22\pm0.33 $\\
$p$        & $\Lambda_c^-$ & $9\pm7 $\\ \hline
\end{tabular}
\end{center}
\end{table}
In general, we observe a very hard production characteristic for all beam
particles. There is a striking contrast in $\Lambda_c^-$ production between
$\pi^-$ beam and the baryon beams.  In addition, we see a structure at large
$x_F$  in the $\Sigma^-$ data which suggests, together with the other 
observations, Pythia--style color drag.

For the production behavior of $\Lambda_c^+$ in $p_T$, 
we observe the same Gaussian slope for $p_T<4\,\mbox{GeV/}c$ for all 3~beam
particles. At higher $p_T$, in the $\Sigma^-$ data we observe a deviation
from the Gaussian behavior.
\ifpreprint

\fi
The production asymmetry results are consistent with other
measurements\cite{wa89asym,e791lamasym}, but, especially in the higher
$x_F$ values and for the $\Sigma^-$ beam, of higher statistics.

\subsection{Production of $D_s$ by $\Sigma^-$ and $\pi^-$}
\ifpreprint\else\looseness=-1\fi
We observe a large difference in the $\Sigma^-$ data for the production
of $D_s^-$ ($n=3.8\pm0.4$) and $D_s^+$ ($n=7.1\pm0.9$), with an integrated
asymmetry (defined as $A\equiv(N_{D_s^-}-N_{D_s^+})/(N_{D_s^-}+N_{D_s^+})$)
of $A=0.57\pm0.07\;(x_F>0.2)$. For the $\pi^-$ data, we obtain 
$A=-0.08\pm0.08$. This favors again the color--drag picture of production.

\subsection{Production of $D^0$ by $\Sigma^-$, $\pi^-$, $p$}
\looseness=-1
We measured the $x_F$ distribution of the $D^0$ and $\overline{D^0}$
production. The preliminary results are shown in table~\ref{d0pmn}.
\begin{table}[htb]
\caption{ \it Preliminary results for a fit to the differential cross section
~~~ $d\sigma/dx_F \sim (1-x_F)^n$ for $D^0$, $\overline{D^0}$
and $D^+$, $D^-$ for different beams.
\label{d0pmn}
}
\vskip 0.1 in
\begin{minipage}{5cm}
\begin{tabular}{|l|c|c|}
\hline
Beam & Particle & $n$ \\ \hline
$\Sigma^-$ & $D^0$            & $6.20\pm0.27$ \\
$\Sigma^-$ & $\overline{D^0}$ & $7.30\pm 0.26 $\\ \hline
$\pi^-$    & $D^0$            & $3.65\pm0.35 $\\
$\pi^-$    & $\overline{D^0}$ & $5.04\pm0.44 $\\ \hline
$p$        & $D^0$            & $5.88\pm0.46 $\\
$p$        & $\overline{D^0}$ & $7.30\pm0.26 $\\ \hline
\end{tabular}
\end{minipage}
\hfill
\begin{minipage}{5cm}
\begin{tabular}{|l|c|c|}
\hline
Beam & Particle & $n$ \\ \hline
$\Sigma^-$ & $D^+$            & $4.95\pm0.23$ \\
$\Sigma^-$ & $D^-$            & $4.67\pm0.22 $\\ \hline
$\pi^-$    & $D^+$            & $2.46\pm0.31 $\\
$\pi^-$    & $D^-$            & $3.58\pm0.34 $\\ \hline
$p$        & $D^+$            & $4.42\pm0.42 $\\
$p$        & $D^-$            & $4.74\pm0.40 $\\ \hline
\end{tabular}
\end{minipage}
\end{table}
\ifpreprint

\fi
A comparison of $\pi^-$ data from E791\cite{e791d0} for $d\sigma/dx_F$
for the sum of of $D^0+\overline{D^0}$ shows good agreement with exception 
at the highest
$x_F$ values, where we do not observe a change in the $(1-x_F)^n$ 
behavior.

\subsection{Production of $D^\pm$ by $\Sigma^-$, $\pi^-$, $p$}
The preliminary results for the $x_F$ behavior of $D^\pm$ production
by the 3~different beams are shown in table~\ref{d0pmn}.
For the $D^\pm$ production asymmetry 
$A\equiv(N_{D^+}-N_{D^-})/(N_{D^+}+N_{D^-})$
as a function of $x_F$ we observe a nearly constant, slightly negative
value of $A$. For the $\pi^-$ data at higher $x_F$ we observe $A$
consistent with $0$, which is, due to the large errors, not totally
inconsistent with previous results\cite{e791dasym,wa92dasym}, but
indicates some problem in one of the measurements.  

\subsection{Summary of Hadroproduction Results}
As a conclusion we like to mention that our data in general
show a strong sensitivity to shared valence quarks. In contrary, they provide
little support for the ``intrinsic charm'' model.

\section{Hyperon Physics}

\subsection{$\Sigma^-$ Radius Measurement}
\looseness=-1
The measurement of the $\Sigma^-$ electromagnetic charge radius was performed
in parallel with our charm data taking.  A special trigger (and a subsequent
analysis) selected events with two and only two outgoing
tracks, one of them an electron, which were supposed to come from a 
beam particle scattered on a target electron.  The differential cross
section for this process is given by
\begin{equation}
{{d\sigma}\over{dQ^2}} = {{4\pi\alpha^2\hbar^2}\over{Q^4}}
\left( 1-{{Q^2}\over{Q_{\rm max}^2}}\right) \cdot F^2(G_E,G_M,Q^2)
\end{equation}
and
\begin{equation}
G_E(Q^2) = {{1}\over{\kappa-1}} G_M(Q^2) = 
\left( 1+{{1}\over{12}}Q^2\!<\!\!r_{ch}^2\!\!>\right)^{-2}
\label{eq2}
\end{equation}
where $Q$ denotes the momentum transfer in the reaction and $Q_{\rm max}$ its
kinematically allowed maximum.  Via approximating the electric and magnetic
form factors $G_E$ and $G_M$ with the dipole approximation (eq.~\ref{eq2}),
the mean squared charge radius $<\!\!r_{ch}^2\!\!>$ can be extracted by a fit
to the slope of the $Q^2$ distribution.

\looseness=-1
We applied this method to $\Sigma^-$, $\pi^-$, and $p$ as beam particle,
the latter two to validate the method. The preliminary results are shown
in table~\ref{radius}.
\begin{table}[htb]
\begin{center}
\caption{ \it Preliminary results for the mean squared charge radius
$<\!\!r_{ch}^2\!\!>$ for different particles compared to previous measurements.
\label{radius}
}
\vskip 0.1 in
\begin{tabular}{|l|c|c|}
\hline
   &SELEX $<\!\!r^2\!\!>[\mbox{fm}^2]$& 
$<\!\!r^2\!\!>[\mbox{fm}^2]$\\ \hline
$\Sigma^-$ & $0.61 \pm 0.12 \pm 0.09$ & 
 $0.91 \pm 0.51$\cite{wa89radius} \\ \hline
$p$        & $0.69 \pm 0.06 \pm 0.06$ & 
 $0.72 \pm 0.01$\cite{mergell} \\ \hline
$\pi^-$    &  $0.42 \pm 0.06 \pm 0.08$ & 
 $0.44 \pm 0.01$\cite{na7}\\ \hline
\end{tabular}
\end{center}
\end{table}
As can be seen, the $\Sigma^-$ result is the first measurement
(after the feasibility has been demonstrated\cite{wa89radius}) of this
type. Our results for $p$ and $\pi^-$ are consistent with other, 
high statistics, measurement.

We extracted also the strong interaction radius of the $\Sigma^-$ from
data of our total cross section measurement\cite{uwe} and obtain a consistent
result.

\subsection{$\Lambda^0$ inclusive polarization}
SELEX made a new measurement of the polarization of inclusive
$\Lambda^0$s produced by $\Sigma^-$, extending to higher $x_F$ than a 
previous measurement\cite{wa89lampol}.
The preliminary results shows an opposite sign, compared to the previous
measurement. A detailed comparison of the analysis is currently in progress.

\subsection{$\Sigma^+$ polarization} 
During a dedicated running period, SELEX took data to measure the polarization
of $\Sigma^+$s produced by protons with a momentum of $800\,\mbox{GeV/}c$.
This measurement was performed at several points in $x_F$ and $p_T$, and
for copper and beryllium targets, extending
previous measurements\cite{e761}, but with lower statistics. 
Even though the errors are large, our date
indicate a slightly higher polarization when the $\Sigma^+$ are produced
in a Beryllium target.

\section{Conclusion}
SELEX submitted the most precise measurement of the $\Lambda_c$ lifetime
for publication: $\tau(\Lambda_c)= (198.1\pm 7.0\pm 5.1)\,\mbox{fs}$.
A preliminary result for the $D_s$ lifetime 
$\tau(D_s^\pm)=(475.6\pm 17.5\pm 4.4)\,\mbox{fs}$
will be submitted soon. The analysis method was validated in both cases
by measuring the $D^0$ and $D^+$ lifetimes.

SELEX has new results on hadroproduction of 
$\Lambda_c$, $D_s$, $D^0$, and $D^\pm$
with $\Sigma^-$, $\pi^-$ and $p$ beams.

In the hyperon sector, SELEX has new results on the electromagnetic
charge radius of the $\Sigma^-$, the polarization of inclusive
$\Lambda^0$ produced by $\Sigma^-$, and the polarization of $\Sigma^+$ 
produced by protons.

We are starting now a second pass over all data.  We improved significantly
our efficiency to reconstruct hyperons, which should help us in accessing
other charm states, especially charm strange baryons.

\section{Acknowledgement}
The authors are indebted to the staff of Fermi National Accelerator Laboratory
and for invaluable technical support from the staffs of collaborating
institutions.
This project was supported in part by Bundesministerium f\"ur Bildung, 
Wissenschaft, Forschung und Technologie, Consejo Nacional de 
Ciencia y Tecnolog\'{\i}a {\nobreak (CONACyT)},
Conselho Nacional de Desenvolvimento Cient\'{\i}fico e Tecnol\'ogico,
Fondo de Apoyo a la Investigaci\'on (UASLP),
Funda\c{c}\~ao de Amparo \`a Pesquisa do Estado de S\~ao Paulo (FAPESP),
the Israel Science Foundation founded by the Israel Academy of Sciences and 
Humanities, Istituto Nazionale di Fisica Nucleare (INFN),
the International Science Foundation (ISF),
the National Science Foundation (Phy \#9602178),
NATO (grant CR6.941058-1360/94),
the Russian Academy of Science,
the Russian Ministry of Science and Technology,
the Turkish Scientific and Technological Research Board (T\"{U}B\.ITAK),
the U.S. Department of Energy (DOE grant DE-FG02-91ER40664 and DOE contract
number DE-AC02-76CHO3000), and
the U.S.-Israel Binational Science Foundation (BSF).


\begin{thebibliography}{99}


\bibitem{spec}   J.~Russ {\it et al.},\ in: Proceedings of the
                 29th International Conference on High Energy Physics,
                 ed.\ A.~Astbury {\it et al.}\, (World Scientific,
                 Singapore, 1999) VolII, p.~1259. 
 \ifpreprint hep--ex/9812031.  \fi

\bibitem{rich}   J.~Engelfried {\it et al.},\ Nucl.\ Instr.\ and Methods 
{\bf A431}, 53 (1999). \ifpreprint hep--ex/9811001. \fi

\bibitem{lambdac} SELEX Collaboration, A.~Kushnirenko {\it et al.},\
submitted to PRL. Preprint Fermilab--Pub--00/255--E, hep--ex/0010014 (2000).


\bibitem{lambdaca} E691 Collaboration, J.C.~Anjos {\it et al.},\ 
Phys.\ Rev.\ Letters {\bf 60} 1379 (1988).

\bibitem{lambdacb} ACCMOR Collaboration, S.~Barlag {\it et al.},\ 
Phys.\ Letters {\bf B218} 374 (1989).

\bibitem{lambdacc} E687 Collaboration, P.L.~Frabetti {\it et al.},\ 
Phys.\ Letters {\bf B251} 639 (1990).

\bibitem{lambdacd} NA14/2 Collaboration, M.P.~Alvarez {\it et al.},\ 
Z.\ Phys.\ {\bf C47} 539 (1990).

\bibitem{lambdace} E687 Collaboration, P.L.~Frabetti {\it et al.},\ 
Phys.\ Rev.\ Letters {\bf 70} 1755 (1993).

\bibitem{cleolc} CLEO Collaboration, A.H.~Mahmood {\it et al.},\
submitted for publication. Preprint CLNS 00/1701, hep--ex/0011049 (2000).

\bibitem{focuslc} FOCUS Collaboration, G.~Boca, talk at this conference (2000).

\bibitem{d0a} E691 Collaboration, J.R.~Raab {\it et al.},\ 
Phys.\ Review {\bf D37} 2391 (1988).

\bibitem{d0b} ACCMOR Collaboration, S.~Barlag {\it et al.},\ 
Z.\ Phys.\ {\bf C46} 563 (1990).

\bibitem{d0c} E687 Collaboration, P.L.~Frabetti {\it et al.},\ 
Phys.\ Letters {\bf B263} 584 (1991).

\bibitem{d0d} E687 Collaboration, P.L.~Frabetti {\it et al.},\ 
Phys.\ Letters {\bf B323} 456 (1994).

\bibitem{d0e} CLEO Collaboration, G.~Bonvicini {\it et al.},\ 
Phys.\ Rev.\ Letters {\bf 82} 4586 (1999).

\bibitem{d0f} E791 Collaboration, E.M.~Aitala {\it et al.},\ 
Phys.\ Rev.\ Letters {\bf 83} 32 (1999).

\bibitem{focusd0} FOCUS Collaboration, S.P.~Ratti, talk at 
``4th Recontres du Vietnam'', Hanoi, Vietnam, 18-25 July, 2000.

\bibitem{dsa} ARGUS Collaboration, H.~Albrecht {\it et al.},\ 
Phys.\ Letters {\bf B210} 267 (1988).

\bibitem{dsb} E687 Collaboration, P.L.~Frabetti {\it et al.},\ 
Phys.\ Rev.\ Letters {\bf 71} 827 (1993).

\bibitem{dsc} E791 Collaboration, E.M.~Aitala {\it et al.},\  
Phys.\ Letters {\bf B445} 449 (1999).

\bibitem{pdg2000} Particle Data Group, D.E.~Groom {\it et al.},\ 
Eur.\ Phys.\ J.\ {\bf C15}, 1 (2000).

\bibitem{pdg1998} Particle Data Group, C.~Caso {\it et al.},\ 
Eur.\ Phys.\ J.\ {\bf C3}, 1 (1998).

\bibitem{bigi} I.I.~Bigi, N.G.~Uraltsev, Z.\ Phys.\ {\bf C62} 623 (1994).

\bibitem{leading} S.~Frixione, M.~Mangano, P.~Nason and G.~Ridolfi
                 ``Heavy quark production'' in ``Heavy Flavour II'', 
                 A.~Buras and M.~Lindner eds.,\ World Scientific
                 Publishing, Singapore (1997). 
  \ifpreprint hep--ph/9702287. \fi

\bibitem{brodski}
S.J.~Brodsky {\it et al.},\ Phys.\ Letters {\bf B93}, 451 (1980)\\
S.J.~Brodsky, C.~Peterson, N.~Sakai, Phys.\ Review {\bf D23}, 2745 (1981)\\
S.J.~Brodsky {\it et al.},\ Nucl.\ Phys.\ {\bf B369}, 519 (1992)\\
R.~Vogt, S.J.~Brodsky, Nucl.\ Phys.\ {\bf B438}, 261 (1995)\\
R.~Vogt, S.J.~Brodsky, Nucl.\ Phys.\ {\bf B478}, 311 (1996).

\bibitem{wa89asym} WA89 Collaboration, M.I.~Adamovich {\it et al.},\ 
Eur.\ Phys.\ J.\ {\bf C8} 593 (1999). \ifpreprint hep--ex/9803021. \fi 

\bibitem{e791lamasym}
 E791 Collaboration, E.M.~Aitala {\it et al.},\ 
submitted for publication, hep--ex/0008029.

\bibitem{e791d0} 
 E791 Collaboration, E.M.~Aitala {\it et al.},\ 
Phys.\ Letters {\bf B462} 225 (1999). 
\ifpreprint hep--ex/9906034. \fi

\bibitem{e791dasym}
 E791 Collaboration, E.M.~Aitala {\it et al.},\ 
Phys.\ Letters {\bf B371} 157 (1996). 

\bibitem{wa92dasym}
WA92 Collaboration, M.I.~Adamovich {\it et al.},\ 
Nucl.\ Phys.\ {\bf B495} 3 (1997).

\bibitem{wa89radius}
WA89 Collaboration, M.I.~Adamovich {\it et al.},\ 
Eur.\ Phys.\ J.\ {\bf C8} 59 (1999).

\bibitem{mergell}
Mergell {\it et al.},\ Nucl.\ Phys.\  {\bf A596} 596 (1996).

\bibitem{na7}
NA7 Collaboration, Nucl.\ Phys.\ {\bf A277} 168 (1986).

\bibitem{uwe}
SELEX Collaboration, U.~Dersch {\it et al.},\ 
Nucl.\ Phys.\ {\bf B579} 277 (2000). \ifpreprint hep--ex/9910052. \fi

\bibitem{wa89lampol}
WA89 Collaboration, M.I.~Adamovich {\it et al.},\ 
Z.\ Phys.\ {\bf A350} 379 (1995). \ifpreprint CERN-PPE/94-86.\fi 

\bibitem{e761}
E761 Collaboration, A.~Morelos {\it et al.},\ 
Phys.\ Rev.\ Letters {\bf 71} 2172 (1993)\\ 
E761 Collaboration, A.~Morelos {\it et al.},\ 
Phys.\ Review {\bf D52} 3777 (1995).

\ifpreprint

\bibitem[$^\ast$]{}deceased
\bibitem[$^\dag$]{}Present address: SAP, Walldorf, Germany
\bibitem[$^\ddag$]{}Now at Imperial College, London SW7 2BZ, U.K.
\bibitem[$^\S$]{}Now at Instituto de F\'{\i}sica da Universidade Estadual de 
Campinas, UNICAMP, SP, Brazil
\bibitem[$^\P$]{}Now at Physik-Department, Technische Universit\"at 
M\"unchen, \\
85748 Garching, Germany
\bibitem[$^\parallel$]{}Present address: The Boston Consulting Group,
M\"unchen, Germany
\bibitem[$^{\ast\ast}$]{}Present address: Lucent Technologies, Naperville, IL
\bibitem[$^{\dag\dag}$]{}Now at Instituto de F\'{\i}sica Te\'orica da 
Universidade Estadual Paulista, S\~ao Paulo, Brazil
\bibitem[$^{\ddag\ddag}$]{}Present address: SPSS Inc., Chicago, IL
\bibitem[$^{\S\S}$]{}Now at University of Alabama at Birmingham, 
Birmingham, AL 35294
\bibitem[$^{\P\P}$]{}Present address: DOE, Germantown, MD
\bibitem[$^{\ast\ast\ast}$]{} Present address: Siemens Medizintechnik, 
Erlangen, Germany
\bibitem[$^{\dag\dag\dag}$]{}Present address: Deutsche Bank AG, 
Eschborn, Germany



\else 
\bibitem{selex} The SELEX Collaboration: 
G.\ Alkhazov$^{k}$,
A.G.\ Atamantchouk$^{k}$,
M.Y.\ Balatz$^{h,}$\footnote{deceased},
\newcounter{marka}\setcounter{marka}{\thefootnote}
N.F.\ Bondar$^{k}$,
P.S.\ Cooper$^{e}$,
L.J.\ Dauwe$^{q}$,
G.V.\ Davidenko$^{h}$,
U.\ Dersch$^{i,}$\footnote{Present address: SAP, Walldorf, Germany},
A.G.\ Dolgolenko$^{h}$,
G.B.\ Dzyubenko$^{h}$,
R.\ Edelstein$^{c}$,
L.\- E\-me\-dia\-to$^{s}$,
A.M.F.\ Endler$^{d}$,
J.\ Engelfried$^{m,e}$,
I.\ Eschrich$^{i,}$\footnote{Now at Imperial College, London SW7 2BZ, U.K.},
C.O.\ Escobar$^{s,}$\footnote{Now at Instituto de F\'{\i}sica da 
Universidade Estadual de Campinas, UNICAMP, SP, Brazil},
A.V.\ Evdokimov$^{h}$,
I.S.\ Filimonov$^{j,}$\footnotemark[\themarka],
F.G.\ Garcia$^{s,e}$,
M.\ Gaspero$^{r}$,
I.\ Giller$^{l}$,
V.L.\ Golovtsov$^{k}$,
P.\ Gouffon$^{s}$,
E.\ G\"ulmez$^{b}$,
He\ Kangling$^{g}$,
M.\ Iori$^{r}$,
S.Y.\ Jun$^{c}$,
M.\ Kaya$^{p}$,
J.\ Kilmer$^{e}$,
V.T.\ Kim$^{k}$,
L.M.\ Kochenda$^{k}$,
I.\ Konorov$^{i,}$\footnote{Now at Physik-Department, Technische Universit\"at 
M\"unchen,\\ 85748 Garching, Germany},
A.P.\ Kozhevnikov$^{f}$,
A.G.\ Krivshich$^{k}$,
H.\ Kr\"uger$^{i,}$\footnote{Present address: The Boston Consulting Group,
M\"unchen, Germany},
M.A.\ Kubantsev$^{h}$,
V.P.\ Kubarovsky$^{f}$,
A.I.\ Kulyavtsev$^{c,}$\footnote{Present address: Lucent Technologies,
Naperville, IL},
\newcounter{markb}\setcounter{markb}{\thefootnote}
N.P.\ Kuropatkin$^{k}$,
V.F.\ Kurshetsov$^{f}$,
A.\ Kushnirenko$^{c}$,
S.\ Kwan$^{e}$,
J.\ Lach$^{e}$,
A.\ Lamberto$^{t}$,
L.G.\ Landsberg$^{f}$,
I.\ Larin$^{h}$,
E.M.\ Leikin$^{j}$,
Li\ Yunshan$^{g}$,
M.\ Luksys$^{n}$,
T.\ Lungov$^{s,}$\footnote{Now at Instituto de F\'{\i}sica Te\'orica da 
Universidade Estadual Paulista, S\~ao Paulo, Brazil},
V.P.\ Maleev$^{k}$,
D.\ Mao$^{c,}$\footnotemark[\themarkb],
Mao\ Chensheng$^{g}$,
Mao\ Zhenlin$^{g}$,
P.\ Mathew$^{c,}$\footnote{Present address: SPSS Inc., Chicago, IL},
M.\ Mattson$^{c}$,
V.\ Matveev$^{h}$,
E.\ McCliment$^{p}$,
M.A.\ Moinester$^{l}$,
V.V.\ Molchanov$^{f}$,
A.\ Morelos$^{m}$,
K.D.\ Nelson$^{p,}$\footnote{Now at University of Alabama at Birmingham,
Birmingham, AL 35294},
A.V.\ Nemitkin$^{j}$,
P.V.\ Neoustroev$^{k}$,
C.\ Newsom$^{p}$,
A.P.\ Nilov$^{h}$,
S.B.\ Nurushev$^{f}$,
A.\ Ocherashvili$^{l}$,
Y.\ Onel$^{p}$,
E.\ Ozel$^{p}$,
S.\ Ozkorucuklu$^{p}$,
A.\ Penzo$^{t}$,
S.V.\ Petrenko$^{f}$,
P.\ Pogodin$^{p}$,
M.\ Procario$^{c,}$\footnote{Present address: DOE, Germantown, MD},
V.A.\ Prutskoi$^{h}$,
E.\ Ramberg$^{e}$,
G.F.\ Rappazzo$^{t}$,
B.V.\ Razmyslovich$^{k}$,
V.I.\ Rud$^{j}$,
J.\ Russ$^{c}$,
P.\ Schiavon$^{t}$,
J.\ Simon$^{i,}$\footnote{Present address: Siemens Medizintechnik, Erlangen, 
Germany},
A.I.\ Sitnikov$^{h}$,
D.\ Skow$^{e}$,
V.J.\ Smith$^{o}$,
M.\ Srivastava$^{s}$,
V.\ Steiner$^{l}$,
V.\ Stepanov$^{k}$,
L.\ Stutte$^{e}$,
M.\ Svoiski$^{k}$,
N.K.\ Terentyev$^{k,c}$,
G.P.\ Thomas$^{a}$,
L.N.\ Uvarov$^{k}$,
A.N.\ Vasiliev$^{f}$,
D.V.\ Vavilov$^{f}$,
V.S.\ Verebryusov$^{h}$,
V.A.\ Victorov$^{f}$,
V.E.\ Vishnyakov$^{h}$,
A.A.\ Vorobyov$^{k}$,
K.\ Vorwalter$^{i,}$\footnote{Present address: Deutsche Bank AG, Eschborn,
Germany},
J.\ You$^{c,e}$,
Zhao\ Wenheng$^{g}$,
Zheng\ Shuchen$^{g}$,
R.\ Zukanovich-Funchal$^{s}$.
$^a$Ball State University, Muncie, IN 47306, U.S.A.
$^b$Bogazici University, Bebek 80815 Istanbul, Turkey.
$^c$Carnegie-Mellon University, Pittsburgh, PA 15213, U.S.A.
$^d$Centro Brasiliero de Pesquisas F\'{\i}sicas, Rio de Janeiro, Brazil.
$^e$Fermilab, Batavia, IL 60510, U.S.A.
$^f$Institute for High Energy Physics, Protvino, Russia.
$^g$Institute of High Energy Physics, Beijing, P.R.\ China.
$^h$Institute of Theoretical and Experimental Physics, Moscow, Russia.
$^i$Max-Planck-Institut f\"ur Kernphysik, 69117 Heidelberg, Germany.
$^j$Moscow State University, Moscow, Russia.
$^k$Petersburg Nuclear Physics Institute, St.\ Petersburg, Russia.
$^l$Tel Aviv University, 69978 Ramat Aviv, Israel.
$^m$Universidad Aut\'onoma de San Luis Potos\'{\i}, San Luis Potos\'{\i}, Mexico.
$^n$Universidade Federal da Para\'{\i}ba, Para\'{\i}ba, Brazil.
$^o$University of Bristol, Bristol BS8~1TL, United Kingdom.
$^p$University of Iowa, Iowa City, IA 52242, U.S.A.
$^q$University of Michigan--Flint, Flint, MI 48502, U.S.A.
$^r$University of Rome ``La Sapienza'' and INFN, Rome, Italy.
$^s$University of S\~ao Paulo, S\~ao Paulo, Brazil.
$^t$University of Trieste and INFN, Trieste, Italy.
\fi


\end{thebibliography}
\end{document}